# Structure, Thermodynamic and Electronic Properties of Carbon-Nitrogen Cubanes and Protonated Polynitrogen Cations

Vitaly V. Chaban[1] and Nadezhda A. Andreeva[2]

1) P.E.S., Vasilievsky Island, Saint Petersburg, Leningrad oblast, Russian Federation.

2) PRAMO, St Petersburg, Leningrad oblast, Russian Federation.

**Abstract**. Energy generation and storage are at the center of modern civilization. Energetic materials constitute quite a large class of compounds with a high amount of stored chemical energy that can be released. We hereby use a combination of quantum chemistry methods to investigate feasibility and properties of carbon-nitrogen cubanes and multi-charged polynitrogen cations in the context of their synthesis and application as unprecedented energetic materials. We show that the stored energy increases gradually with the nitrogen content increase. Nitrogen-poor cubanes retain their stabilities in vacuum, even at elevated temperatures. Such molecules will be probably synthesized at some point. In turn, polynitrogen cations are highly unstable, except N8H+, despite they are isoelectronic to all-carbon cubane. Kinetic stability of the cation decays drastically as its total charge increases. High-level thermodynamic calculations revealed that large amounts of energy are liberated upon decompositions of polynitrogen cations, which produce molecular nitrogen, acetylene, and protons. The present results bring a substantial insights to the design of novel high-energy compounds.

**TOC Image**

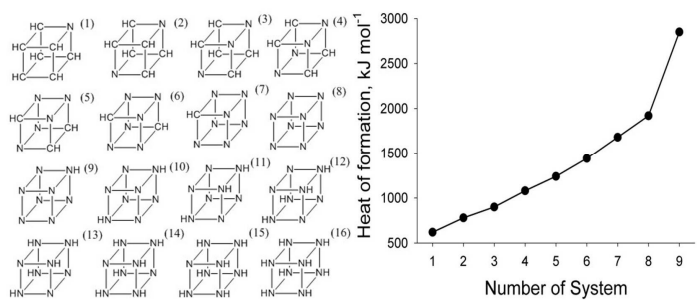

**Introduction**

Due to its ability to maintain stable chains of homonuclear covalent bonds, nitrogen attracts a substantial interest of chemical community.[1-14] Nitrogen exhibits a diversity of phases in the solid state.[15-18] Nitrogen-rich molecules and particularly polynitrogen compounds may constitute interest for energy storage, since their decomposition (assuming formation of dinitrogen as a final product) liberates a huge amount of energy. Furthermore, such reactions are environmentally friendly giving rise to green propellants and explosives. While many exotic polynitrogen molecules were predicted by quantum chemical simulations, their synthesis remains challenging and their stability at finite temperatures was demonstrated for a limited number of species only.

High pressures, over $1\times10^3$ bar, foster formation of polynitrogen structures by helping to overcome the energy barrier due to the lone electron pair repulsion. While stable at the conditions of their synthesis, many such nitrogen rich products decompose immediately after the pressure in the reaction vessel is lowered. First-principles thermodynamic calculations are helpful to predict behavior of exotic molecules at extreme conditions. Under pressures exceeding $1.1\times10^3$ bar and temperatures exceeding 2000 K, nitrogen forms a solid network, in which all atoms are connected by a single covalent bond, 0.135 nm long, a cubic-gauche structure with eight nitrogen atoms per unit cell.[19] The cubic-gauche structure ruins spontaneously after the pressure is lowered, being considered the highest-energy, 33 kJ g$^{-1}$, non-nuclear material suitable for explosives and rocket fuels.[3]

A principal possibility of existence of small polynitrogen molecules – such as $N_3^-$ (azide anion), $N_4$ (tetranitrogen), and $N_5$ (pentanitrogen) – was demonstrated.[20,21] Cations $N_4^+$ and $N_5^+$ are considered to be even more stable than the corresponding neutral particles. Such cations can be obtained by an aggressive electron bombardment of $N_2$. Although obtained experimentally, none of the above molecules was properly characterized. Little is known about $N_4$, besides its lifetime being on the order of 1 μs at room conditions.

A number of isomers are possible for $N_n$. For instance, $N_4$ exists both as a linear molecule and as a cycle.[22,23] Computer simulations based on quantum chemical principles are of paramount importance to investigate structure, stability at finite temperature, energy levels, spectroscopic properties, and electronic density distribution. Glukhovtsev and coworkers[24] employed density functional theory (DFT) to suggest applications of polynitrogen compounds as a prospective high energy density matter. A new nitrogen allotrope, $N_6$, was recently predicted, whose structure is an open chain.[25] Due to the large energy difference between the single N-N bond and the triple N≡N bond, dissociation the $N_6$ crystal releases large amounts of energy. Formally, $N_6$ is among the most efficient energy materials. Another theoretically revealed allotrope of nitrogen is all-nitrogen cubane (N-cubane) $N_8$.[26,27] In turn, cubane consisting of carbon atoms (C-cubane) was reported in 1964 by Eaton and Cole, who proved its kinetic stability and synthesized a few derivatives.[28-30] Cubane is frequently referred to as "impossible chemistry" due to unusual arrangements of atoms, positive enthalpies of formation, and sophisticated synthesis pathways. Experimentally prepared explosive derivatives of cubane – octanitrocubane and heptanitrocubane – store 4145 kJ mol$^{-1}$ and 4346 kJ mol$^{-1}$, respectively.[31]

Whereas C-cubane and N-cubane were considered before, a mixed carbon-nitrogen cubane (CN-cubane) $(CH)_{8-n}N_n$ received a very limited attention thus far.[see ref3] Such molecules are expected to be somewhat more stable than $N_8$, but also more energetic than $C_8H_8$. Engelke[32,33] employed Hartree-Fock for geometry optimization and MP2 calculations for total energies to characterize all possible contents and symmetries of CN-cubanes. All CN-cubanes were predicted to be stable molecules, although they exhibit large positive heats of formation. Engelke concludes that the primary reason of the high-energy content of the nitrogen-rich CN-cubanes is not due to bond strain, but thanks to a single nitrogen-nitrogen bond. $C_{8-n}N_n$ molecules, with n > 4, if ever synthesized, must become important energetic materials offering huge capacities of chemical energy. Unfortunately, no investigation of thermal stability of either N-cubane or CN-cubanes has been performed yet. Furthermore, no convincing evidence exists on the decomposition pathways of CN-cubanes and their products at the conditions of degradation.

Observation of how stability of C-cubane at finite temperature decreases in response to a stepwise addition of the nitrogen atoms and, therefore, formation of the carbon-nitrogen polar covalent bonds is of significant fundamental interest. We hereby employ a combination of quantum-chemical methods (composite thermochemistry G4MP2, hybrid density functional theory B3LYP, MP2, and large-timescale semiempirical molecular dynamics PM7-MD) to shed light on the most energetically rich CN-cubanes and their thermal stabilities at room and elevated temperatures. A systematic investigation of $(CH)_{8-n}N_n$ molecules, $1 > n > 8$, and protonated N-cubanes $N_8H_n^{n+}$, $1 > n > 8$. (Figure 1) was performed. The protonated N-cubanes constitute research interest for two reasons: (1) they are isoelectronic to C-cubane that might increase their stabilities; (2) electrostatic repulsion in the n > 1 cases influences stabilities unfavorably.

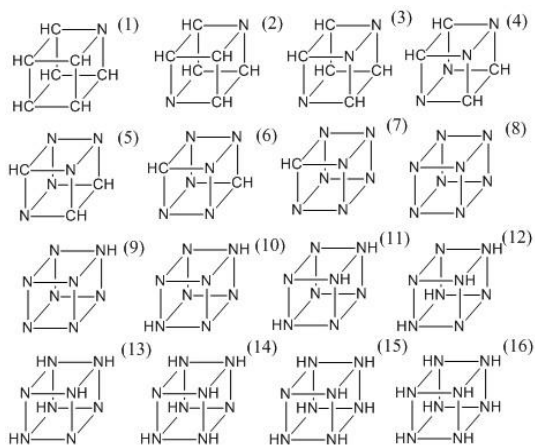

Figure 1. Carbon-nitrogen cubane molecules (1 - 8) and cations obtained via protonation of all-nitrogen cubane (9 - 16): (1) $(CH)_7N$, (2) $(CH)_6N_2$, (3) $(CH)_5N_3$, (4) $(CH)_4N_4$, (5) $(CH)_3N_5$, (6) $(CH)_2N_6$, (7) $(CH)N_7$, (8) $N_8$, (9) $N_8H^+$, (10) $N_8H_2^{2+}$, (11) $N_8H_3^{3+}$, (12) $N_8H_4^{4+}$, (13) $N_8H_5^{5+}$, (14) $N_8H_6^{6+}$, (15) $N_8H_7^{7+}$, (16) $N_8H_8^{8+}$.

**Methodology**

Stabilities of neutral CN-cubanes and ionic N-cubanes at finite temperatures were investigated by PM7-MD. The equations-of-motion were propagated according to the Verlet integration scheme with a time-step of 0.5 fs during 2.5 ns or until the simulated molecule (ion) was decomposed. The temperature was maintained constant by the Andersen thermostat,[34] based on the reassignment of a chosen atom velocity. The new linear velocity is given by Maxwell–Boltzmann statistics for the specified temperature (300, 1000, 1500, 2000 K). We used the unrestricted PM7 Hamiltonian[35-38] to derive immediate forces at every time-step. PM7 was parameterized using experimental and high-level ab initio reference data, augmented by a new type of reference data intended to better define the structure of parameter space.[38] PM7 uses several empirical corrections, such as those for dispersion attraction, hydrogen bonding, peptide bonding, etc.[38-41] The wave function convergence criterion was set to $10^{-6}$ Hartree in the self-consistent-field procedure. Five independent PM7-MD simulations with arbitrary different starting conditions were carried out, as described above, to increase sampling of the phase space.

Thermodynamic potentials (enthalpy, entropy, Gibbs free energy) were derived from the molecular partition functions. The Gaussian-4 composite technique[42] was used. In Gaussian-4, a sequence of ab initio electronic-structure calculations is performed to provide an accurate total energy for a given molecular species. The ideal gas behavior of the simulated species are assumed. No electronic excitations are accounted for, i.e. the ground electronic state is assumed. An accuracy of Gaussian-4 was estimated using a comprehensive benchmarking set of experimental data (a few hundred compounds) and only very modest discrepancies with the experimental data were detected.[ref42??]

The dipole moment was computed from the corresponding self-consistent wave functions for the geometries optimized at the hybrid density functional theory B3LYP/6-31G* level of theory.[43,44] The covalent and non-covalent geometrical parameters were compared using PM7,[38]

B3LYP,[43,44] and MP2[45] using the 6-31+G* and 6-311+G* atom-centered split-valence double- and triple-zeta polarized basis sets.

Starting geometry of the systems were prepared in Gabedit[46] and the output geometries were visualized in VMD.[47] GAMESS (ver. 12/2014)[48] and MOPAC 2012 (openmopac.net) were used for electronic-structure calculations, whereas the thermal motion was simulated by the in-home code.

**Results and Discussion**

Kinetic stabilities of the CN-cubanes and protonated N-cubanes were assessed at 300, 1000, 1500, and 2000 K (Table 1). The molecular geometries were preliminarily heated to the requested temperature within 10 ps by gradually increasing the reference temperature of the Andersen thermostat. Heating of single molecules occurs uniformly, because energy exchange within a covalently bound structure is extremely quick. After heating, conventional equilibrium molecular dynamics simulations were conducted during up to 2.5 ns. All CN-cubanes $(CH)_{8-n}N_n$ are stable up to 1000 K. This is in line with the conclusions of the pioneering work by Engelke who used restricted Hatree-Fock calculation using a number of isomers.[32,33] At 1500 K, the carbon-rich molecules, e.g. $(CH)_6N_2$ and $(CH)_5N_3$, are generally more stable than nitrogen-rich molecules, e.g. $(CH)_2N_6$ and $(CH)N_7$. Interestingly, $(CH)_7N$ decomposes upon heating to 1500 K, despite containing only one nitrogen atom. Nitrogen, which is more electronegative than carbon, shifts electronic density towards itself and, therefore, ruins initial symmetry of C-cubane. The resulting electronically polarized structure is relatively unstable, as compared to $(CH)_6N_2$, $(CH)_5N_3$, and $(CH)_3N_5$, in which nitrogen atoms are distributed more uniformly. The cationic N-cubanes $N_{8-n}H_n^{n+}$ are systematically less thermally stable than the molecules $(CH)_{8-n}N_n$ containing the same number of nitrogen atoms. Furthermore, larger positive charges undermine stability. This trend is clearly linked to electrostatic repulsion between the atoms belonging to the same cation. $N_8H_2^{2+}$ and $N_8H_3^{3+}$ are stable only at room temperature. All larger charges, except $N_8H_7^{7+}$ and $N_8H_8^{8+}$, are stable at the extremely low temperatures only, whereas their heating to room temperature results in the picosecond-scale liberation of protons $H^+$. The remainder is neutral N-cubane.

Table 1. Stability data for the simulated CN-cubanes and protonated N-cubane cations at 300, 1000, 1500, and 2000 K under high vacuum.

| Compounds | 300 K | 1000 K | 1500 K | 2000 K |
|---|---|---|---|---|
| $(CH)_7N$ | stable | stable | unstable | unstable |
| $(CH)_6N_2$ | stable | stable | stable | unstable |
| $(CH)_5N_3$ | stable | stable | stable | unstable |
| $(CH)_4N_4$ | stable | stable | unstable | unstable |
| $(CH)_3N_5$ | stable | stable | stable | unstable |
| $(CH)_2N_6$ | stable | stable | unstable | unstable |
| $(CH)N_7$ | stable | stable | unstable | unstable |
| $N_8$ | stable | stable | stable | unstable |
| $N_8H^+$ | stable | stable | unstable | unstable |
| $N_8H_2^{2+}$ | stable | unstable | unstable | unstable |
| $N_8H_3^{3+}$ | stable | unstable | unstable | unstable |
| $N_8H_4^{4+}$ | unstable | unstable | unstable | unstable |
| $N_8H_5^{5+}$ | unstable | unstable | unstable | unstable |
| $N_8H_6^{6+}$ | unstable | unstable | unstable | unstable |
| $N_8H_7^{7+}$ * | unstable | unstable | unstable | unstable |
| $N_8H_8^{8+}$ * | unstable | unstable | unstable | unstable |

* The marked ensembles of atoms do not have any local minimum corresponding to the cubic geometry. This result was also confirmed by the electron-correlation local-minimum search at the MP2/6-311+G* level of theory.

Note that our PM7-MD simulations cover only the nanosecond timescale ($5\times10^6$ iterations). Although this duration is enough for all probable chemical transformations to occur, the simulations may not fully account for the so-called rare events. These rare events may launch decomposition of the molecule and may require much longer times, e.g. hours, of real-time sampling. Such long simulations are currently inaccessible, and therefore our conclusions, strictly speaking, apply to sub-nanosecond times only. In addition, our simulations do not account for the bi-molecular decomposition pathways and ignore chemical environment. This is an ideal gas approximation, which is reasonable for small and volatile molecules investigated under high vacuum. It would be a correct supposition that the reported stabilities are systematically overestimated. They are fully valid and informative to compare CN-cubanes and protonated N-cubanes to one another, but should be corrected prior to point-by-point comparison with any future experimental data.

Heats of formation computed at 1000 K (Figure 1) are positive for all CN-cubanes and $N_8H^+$ (the only stable cation in the PM7-MD simulations at 1000 K). It means that CN-cubanes and $N_8H^+$ cannot be directly obtained from simple substances, i.e. molecular nitrogen, molecular hydrogen, and graphite. They may, however, emerge from other nitrogen sources, e.g. thanks to isomerization of polynitrogen compounds at extreme conditions. Substitution of the CH groups by the nitrogen atoms leads to a gradual increase of heat of formation, 640 to 1900 kJ mol$^{-1}$. Thus, the carbon-rich CN-cubanes are more probable than the nitrogen-rich CN-cubanes (Table 1). The $N_8H_1^+$ cation is even less thermodynamically favorable, $\Delta H_f$ = 2850 kJ mol$^{-1}$, but once formed contains the largest portion of energy. The protonated N-cubane cations with larger charges are unstable at 1000 K.

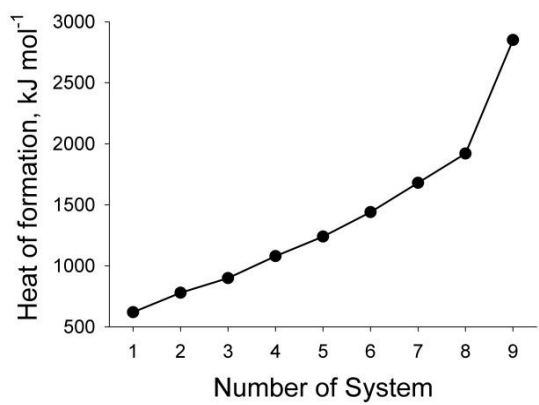

Figure 1. Average heats of formation of all stable molecular and ionic cubanes at 1000 K. Error bars, which were derived based on the thermal fluctuations of this property after equilibration, are smaller than the size of the depicted symbols.

The dipole moments of the CN-cubanes $(CH)_{(8-n)}N_n$ are determined by the locations of the nitrogen atoms (Figure 2). If the molecular structure is symmetric (systems #2, 4, 6, 8), the dipole moment of the relaxed geometry amounts to zero. If the molecular structure is asymmetric, the dipole moment increases from 1.82 to 2.45 D as the number of the nitrogen atoms increases. Relatively small dipole moments suggest that prevailing intermolecular interactions in the condensed state of CN-cubanes are van der Waals, rather than electrostatic ones. Consequently, we expect cohesive energies, and therefore melting and boiling points, in this family of compounds to be quite small, comparable with other molecules of similar masses and containing polar covalent bonds.

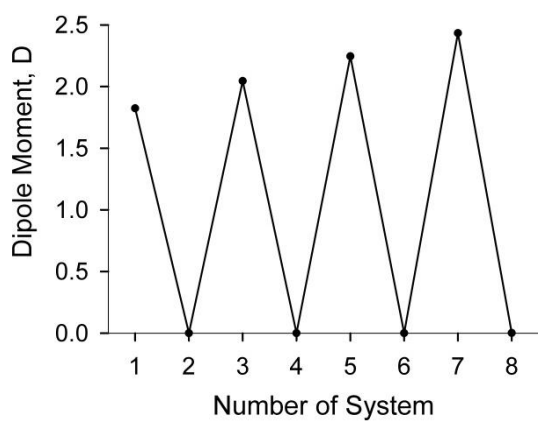

Figure 2. Dipole moments of the $(CH)_{(8-n)}N_n$ cubanes, in which 1 > n > 8. The computations were carried out using tightly equilibrated geometries at the B3LYP/6-31G* level of theory.

We observed in PM7-MD that CN-cubanes decompose with a formation of dinitrogen molecules and acetylene. This is in contrast to decomposition of C-cubane at similar conditions (elevated temperatures and small pressure).[31,49] In particular, C-cubane first isomerizes into cyclooctatetraene $C_8H_8$, which subsequently decomposes into acetylene and benzene. No intermediate formation of heterocycles was observed during decomposition of CN-cubanes. When temperature reaches a critical value, the carbon-nitrogen bonds of the cubic molecule break apart producing a few radicals CH-N*. These radicals readily react in the gas phase resulting in $N_2$ and $C_2H_2$. In no case, we observed benzene, although its formation might have been hypothesized as a by-product. It must be noted that products depend on the content of the atmosphere. For instance, availability of oxygen at high temperatures would result in the further oxidation of $C_2H_2$. To recapitulate, decomposition of the CN-cubanes follows a different pathway, as compared to C-cubane. This important finding must be accounted for during computation of the corresponding thermodynamic potentials.

The role of entropy change (T∆S factor) upon the decomposition of the CN-cubanes is small, -142…-150 kJ mol$^{-1}$, as compared to enthalpy (Figure 3). It is understandable, because these reactions involve breakage of a number of covalent bonds. The decomposition enthalpy decreases significantly in response to each new nitrogen atom substituting the CH group, from +149 in (CH)N to -1755 kJ mol$^{-1}$ in $N_8$. Thus, the carbon-nitrogen polar covalent bonds undermine thermodynamic stability. The free energy decreases in line with enthalpy, from -2 kJ mol$^{-1}$ in $(CH)_7N$ to -1896 kJ mol$^{-1}$ in $N_8$. While $N_8$ was stable at room conditions in the PM7-MD simulations during sub-nanosecond time, its decomposition would liberate nearly 2 MJ mol$^{-1}$ more energy than that of $(CH)_7N$.

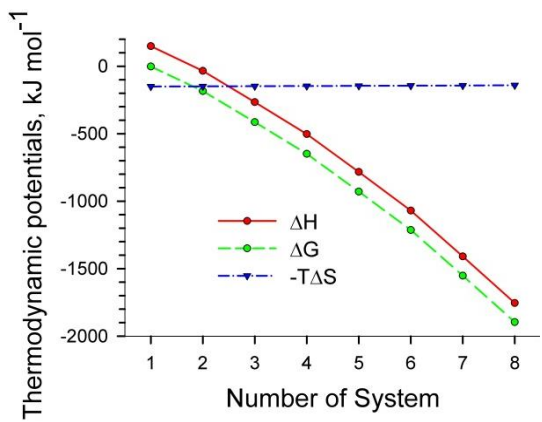

Figure 3. Standard thermodynamic potentials for decomposition of $(CH)_{(8-n)}N_n$, in which 1 < n < 8, into $N_2$ and $C_2H_2$.

Direct comparison of C-cubane and CN-cubanes is not possible, since they produce different products upon decomposition. In particular, decomposition of $(CH)_7N$ liberates only 2 kJ mol$^{-1}$, whereas decomposition of C-cubane onto $C_2H_2$ and $C_6H_6$ brings 353 kJ mol$^{-1}$ (in terms of standard Gibbs free energy), suggesting that $(CH)_7N$ stores less energy. To further corroborate this interesting observation, Figure 4 investigates decomposition of C-cubane. If C-cubane produces 4 $C_2H_2$ molecules, the reaction costs 156 kJ mol$^{-1}$ at room conditions, being therefore thermodynamically forbidden. However, as temperature increases, this reaction decreases its Gibbs free energy drastically, becoming the most preferable decomposition pathway at 2000 K and above. Interestingly, all stages and pathways of the C-cubane decomposition are more thermodynamically favorable at higher temperatures. Note that C-cubane is experimentally known to be perfectly stable at room temperature and starts slow decomposition only above 400 K.

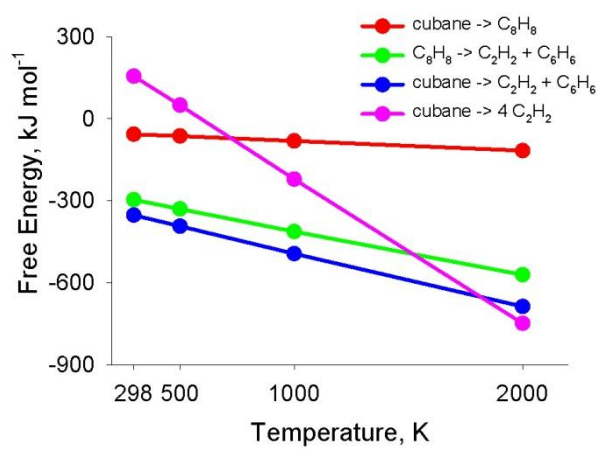

Figure 4. Free energies of C-cubane decomposition at atmospheric pressure and designated temperatures following different pathways and stages: isomerization of C-cubane onto cyclooctatetraene $C_8H_8$ (red line), decomposition of $C_8H_8$ onto acetylene $C_2H_2$ and benzene $C_6H_6$ (green line), direct decomposition of C-cubane onto $C_2H_2$ and $C_6H_6$ (blue line), direct decomposition of C-cubane onto four $C_2H_2$ molecules (pink line).

Cations based on N-cubane $NH_n^{n+}$ are isoelectronic to C-cubane. For this reason, we hypothesize that some of them may appear relatively stable and can exist at low temperatures. Interestingly, $N_8H_2^{2+}$ is slightly more thermodynamically stable than $N_8H^+$ (Figure 5). $N_8H_2^{2+}$ is symmetric, since both hydrogen atoms are located along the same diagonal of the cube. This is, however, not confirmed by PM7-MD simulations at 1000 K, which show that $N_8H^+$ is stable, but $N_8H_2^{2+}$ is not. Thermal motion of the cation violates an initial symmetry. At 300 K, both cations are stable. As expected, further increase of charge undermines stabilities of the cations. $N_8H_7^{7+}$ and $H_8H_8^{8+}$ cannot exist in principle. Very high decomposition enthalpies for the other $N_8H_n^{n+}$ compounds are promising, provided that such cations can ever be obtained.

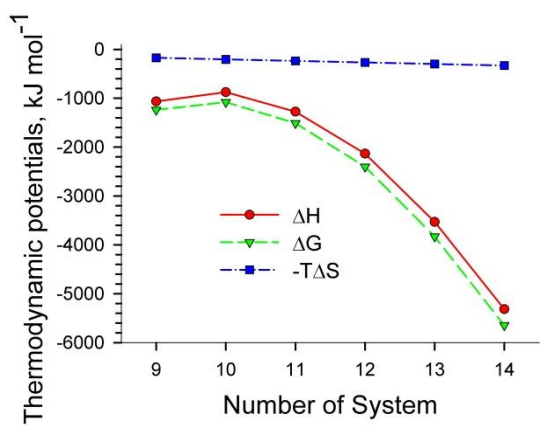

Figure 5. Standard thermodynamic potentials for decomposition of $N_8H_n^{n+}$, in which $1 < n < 6$, $n = 1…6$ into $N_2$ and $H^+$. Systems 15 and 16 are not depicted, since they are unstable, i.e. no local minima correspond to the cubic geometries of such chemical compositions.

Molecular dynamics simulations allow to compute pair correlation functions (Figures 6-8) for the atomic positions, which are useful to characterize molecular geometries in a systematic way. All pair correlation functions contain three peaks, which correspond to the covalent bond distance, the face diagonal, and the space diagonal. Since carbon and nitrogen atoms possess very similar sizes, they can smoothly coexist in the cubic structure, such as CN-cubanes. The carbon-carbon non-polar covalent bond length amounts to 0.158 nm and marginally depends on the number of nitrogen atoms in the structure. This bond length is substantially longer than in most carbonaceous materials, e.g. in graphene and nanotubes, 0.141 nm, since the latter are aromatic. The carbon-nitrogen polar covalent bond length amounts to 0.156 nm, therefore, introduction of nitrogen atoms into the initial C-cubane structure brings a very modest structure perturbation. The nitrogen-nitrogen non-polar covalent bond length is 0.150 nm. Different lengths of the covalent bonds violate symmetry and decrease thermal stability substantially. As exemplified previously,[8] nitrogen-nitrogen bonds undermine thermal stability of the N-doped graphene, but N-doped graphene without the nitrogen-nitrogen bonds is very stable. In the case of CN-cubanes, a similar effect can also be noticed in the nitrogen-rich species (Table 1).

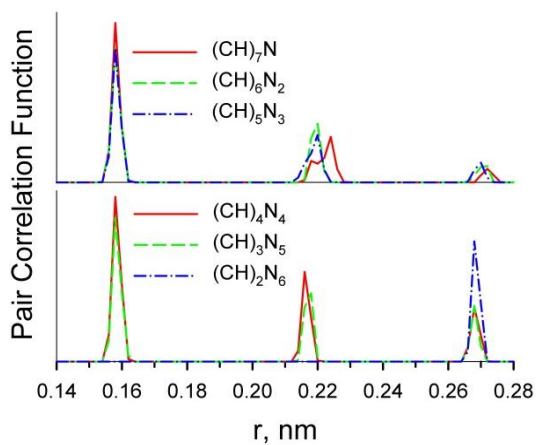

Figure 6. Pair correlation functions computed for carbon-carbon distances in $(CH)_{(8-n)}N_n$, where n = 1…6 at 1000 K.

The face diagonal lengths in CN-cubanes range 0.216 to 0.224 nm, whereas the space diagonal lengths range 0.266 to 0.276 nm at 1000 K. Large fractions of the nitrogen atoms favor somewhat shorter diagonals. If the molecule retains stability at 1500 K, no elongation of diagonals is observed for it, as compared to 1000 K. The nitrogen-nitrogen bonds exhibit the same length in neutral CN-cubanes and stable $NH_n^{n+}$ cations at room temperature.

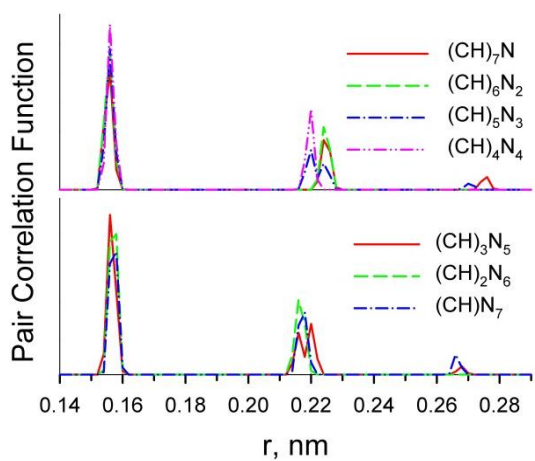

Figure 7. Pair correlation functions computed for carbon-nitrogen distances in $(CH)_{(8-n)}N_n$, in which 1 < n < 7, at 1000 K.

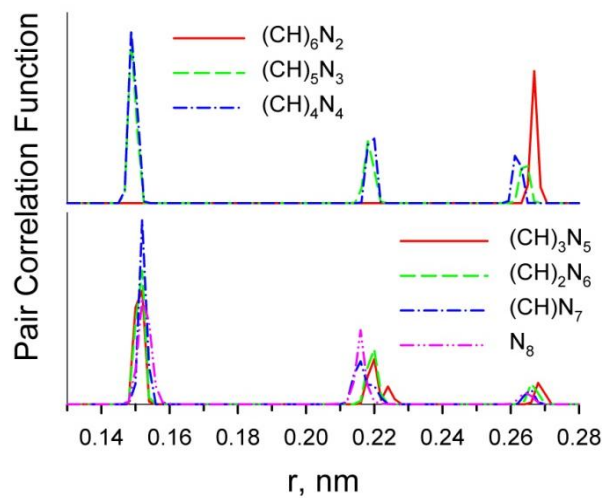

Figure 8. Pair correlation functions computed for nitrogen atoms in $(CH)_{(8-n)}N_n$, where n = 2…8 at 1000 K.

In the remaining part of the paper, we compare molecular structures obtained by PM7 to those obtained by geometry optimizations using post-Hartree-Fock (MP2) and hybrid density functional theory (B3LYP). Both MP2 and B3LYP are very well-established methods with more than two decades of vigorous usage in different fields of chemistry and materials science. Table 2 compares distances and angles for N-cubane $N_8$ employing two basis sets. B3LYP/6-31+G* tends to overestimate distances somewhat. This problem disappears as soon the basis set is converted to triple-zeta one, 6-311+G*. MP2 appears less sensitive to the choice of the basis set, although it is widely believed in the community that DFT is prone to the errors connected with a limited, or generally finite, basis set. Interestingly, B3LYP/6-31G* is more successful in predicting the nitrogen-nitrogen bond length (Table 3) than B3LYP/6-31+G* (Table 2), the only difference between those levels of theory being diffuse basis functions. All methods correctly predict the local-minimum state of $N_8$. Comparison of the geometries obtained by PM7 and four ab initio levels of theory identifies that the results of PM7 are closest to those of the highest employed level of theory, 6-311+G*. Thus, it is once again confirmed that PM7 provides very accurate geometries, in particular for bonded interactions. Thanks to the ability of PM7 to accurately reproduce geometries and heats of formation of a wide variety of molecules, it can be used as a force field substitution in PM7-MD simulations.

Table 2. Point-by-point comparison of molecular geometries obtained by post-Hartree-Fock electron-correlation method (MP2), hybrid density functional theory (B3LYP), and semiempirical electronic-structure method (PM7) as applied to the $N_8$. The highlighted numbers are those from the employed ab initio methods, which are closest to the corresponding result of PM7.

| Covalent bond / Angle | Bond length, nm / Angle, degrees | | | | |
|---|---|---|---|---|---|
| | MP2 6-31+G* | B3LYP 6-31+G* | MP2 6-311+G* | B3LYP 6-311+G* | PM7 |
| N-N (adjacent) | 0.153 | 0.160 | **0.152** | 0.152 | 0.152 |
| N-N (2D diagonal) | 0.217 | 0.227 | **0.215** | 0.214 | 0.215 |
| N-N (3D diagonal) | 0.266 | 0.278 | **0.263** | 0.263 | 0.264 |
| Angle N-N-N | 90 | 90 | **90** | 90 | 90 |

Table 3 summarizes covalent bond lengths in all stable cubic molecules, which were hereby investigated, obtained at the B3LYP/6-31G* level of theory. The computed distances are in perfect agreement with pair correlation functions for nitrogen-nitrogen, carbon-nitrogen, and carbon-carbon distances computed from PM7-MD simulations at 1000 K. Note that the B3LYP/6-31G* local-minimum structures do not include thermal motion and, therefore, the corresponding distances may be slightly smaller than those from the finite-temperature simulations. Overall, a point-by-point comparison of B3LYP and PM7 for the entire set of the simulated molecules and ions reveals very similar geometries, the differences of which are systematically inferior to the thermally induced fluctuations (Figures 6-8).

Table 3. Carbon-carbon, carbon-nitrogen, and nitrogen-nitrogen covalent bonds lengths in all stable molecular CN-cubanes and protonated N-cubane ions obtained at the B3LYP/6-31G* level of theory.

| System | $R_{N-N}$, nm | $R_{C-C}$, nm | $R_{C-N}$, nm | $R_{N-H}$, nm |
|---|---|---|---|---|
| $(CH)_7N$ | — | 0.156…0.158 | 0.153 | — |
| $(CH)_6N_2$ | — | 0.155 | 0.154 | — |
| $(CH)_5N_3$ | 0.156 | 0.153…0.155 | 0.151…0.153 | — |
| (CH)$_4$N$_4$ | 0.157 | 0.152 | 0.152 | — |
| (CH)$_3$N$_5$ | 0.154…0.156 | 0.152 | 0.150…0.152 | — |
| (CH)$_2$N$_6$ | 0.154 | — | 0.149 | — |
| (CH)N$_7$ | 0.151…0.153 | — | 0.150 | — |
| N$_8$ | 0.152 | — | — | — |
| N$_8$H$^+$ | 0.149…0.154 | — | — | 0.104 |
| N$_8$H$_2^{2+}$ | 0.150…0.155 | — | — | 0.105 |
| N$_8$H$_3^{3+}$ | 0.147…0.159 | — | — | 0.107...0.108 |
| N$_8$H$_4^{4+}$ | 0.149…0.161 | — | — | 0.110 |
| N$_8$H$_5^{5+}$ | 0.151…0.166 | — | — | 0.115 |
| N$_8$H$_6^{6+}$ | 0.155…0.164 | — | — | 0.123 |

**Conclusions**

We reported a systematic investigation of the structure, thermodynamic, and electronic properties of the CN-cubanes and multi-charged N-cubane cations. These compounds are hypothetical yet, and computer simulations constitute the only means to investigate their kinetic and thermodynamic stabilities as well as physicochemical properties. The generated knowledge is necessary to understand viability of these compounds. Knowledge of the decomposition mechanisms, in some cases, may help to hypothesize reaction mechanisms.

We found that the energetic capacity of the CN-cubanes increases with the number of the nitrogen atoms in their structure, reaching 1900 kJ mol$^{-1}$ for N$_8$. Therefore, the CN-cubanes are even more energy-rich containers, as compared to C-cubane. The cubic protonated polynitrogen cations exhibit stabilities in the inverse proportion to their total charges. Under vacuum, N$_8$H$_1^{1+}$ is stable at 1000 K, whereas N$_8$H$_6^{2+}$ and N$_8$H$_6^{3+}$ are stable only at 300 K. N$_8$H$_7^{7+}$ and N$_8$H$_8^{8+}$ are unstable in principle, but +4, +5, and +6 charges are stable at very low temperatures, $\leq 100$ K, over sub-nanosecond time. Spontaneous decomposition of N$_8$H$_6^{6+}$ liberates 5315 kJ mol$^{-1}$ providing four molecules of nitrogen and six free protons. It must be understood that stabilities of any molecules in vacuum are much higher than stabilities in real atmospheric environments, so the computed decomposition temperatures will have to be decreased accordingly.

To recapitulate, combined usage of the electronic-structure methods allows to observe "impossible" and non-existing compounds, estimate a set of their properties and possibly hypothesize possibilities of future syntheses. While formation of the nitrogen-rich CN-cubanes and protonated N-cubane cations is very unlikely even at the extremely conditions (e.g. high pressure), $(CH)_7N$ and $(CH)_6N_2$ may be within reach. Such information may be of substantial assistance to the ongoing molecular design efforts.

## Author Information

E-mail for correspondence: vvchaban@gmail.com (V.V.C.)

## References

(1) Dixon, D. A.; Feller, D.; Christe, K. O.; Wilson, W. W.; Vij, A.; Vij, V.; Jenkins, H. D. B.; Olson, R. M.; Gordon, M. S. Enthalpies of Formation of Gas-Phase N3, N3-, N5+, and N5-from Ab Initio Molecular Orbital Theory, Stability Predictions for N5+N3-and N5+N5-, and Experimental Evidence for the Instability of N5+N3. *J Am Chem Soc* 2004, *126*, 834-843.
(2) Nguyen, M. T. Polynitrogen compounds. *Coordin Chem Rev* 2003, *244*, 93-113.
(3) Zarko, V. E. Searching for ways to create energetic materials based on polynitrogen compounds (review). *Combustion, Explosion, and Shock Waves* 2010, *46*, 121-131.
(4) Chung, G.; Schmidt, M. W.; Gordon, M. S. An Ab Initio Study of Potential Energy Surfaces for N8Isomers. *The Journal of Physical Chemistry A* 2000, *104*, 5647-5650.
(5) Nguyen, M. T.; Ha, T.-K. Decomposition mechanism of the polynitrogen N5 and N6 clusters and their ions. *Chem Phys Lett* 2001, *335*, 311-320.
(6) Ponec, R.; Roithová, J.; Gironés, X.; Jug, K. On the nature of bonding in N5+ ion. *Journal of Molecular Structure: THEOCHEM* 2001, *545*, 255-264.
(7) Dang, Q. Q.; Zhan, Y. F.; Wang, X. M.; Zhang, X. M. Heptazine-Based Porous Framework for Selective CO2 Sorption and Organocatalytic Performances. *Acs Appl Mater Inter* 2015, *7*, 28452-28458.
(8) Chaban, V. V.; Prezhdo, O. V. Nitrogen-Nitrogen Bonds Undermine Stability of N-Doped Graphene. *J Am Chem Soc* 2015, *137*, 11688-11694.
(9) Zhang, X.; Yang, J.; Lu, M.; Gong, X. Structure, stability and intramolecular interaction of M(N5)2 (M = Mg, Ca, Sr and Ba) : a theoretical study. *RSC Adv* 2015, *5*, 21823-21830.
(10) Noyman, M.; Zilberg, S.; Haas, Y. Stability of Polynitrogen Compounds: The Importance of Separating the σ and π Electron Systems†. *The Journal of Physical Chemistry A* 2009, *113*, 7376-7382.
(11) Lian, P.; Lai, W.; Chang, H.; Li, Y.; Li, H.; Yang, W.; Wang, Y.; Wang, B.; Xue, Y. Density Functional Theoretical Study of Polynitrogen Compounds N5+Y– (Y=B(CF3)4, BF4, PF6 and B(N3)4). *Chin. J. Chem .* 2012, *30*, 639-643.
(12) Vij, A.; Wilson, W. W.; Vij, V.; Tham, F. S.; Sheehy, J. A.; Christe, K. O. Polynitrogen Chemistry. Synthesis, Characterization, and Crystal Structure of Surprisingly Stable Fluoroantimonate Salts of N5+. *J Am Chem Soc* 2001, *123*, 6308-6313.
(13) Östmark, H.; Wallin, S.; Brinck, T.; Carlqvist, P.; Claridge, R.; Hedlund, E.; Yudina, L. Detection of pentazolate anion (cyclo-N5–) from laser ionization and decomposition of solid p-dimethylaminophenylpentazole. *Chem Phys Lett* 2003, *379*, 539-546.

(14)    Liang, Y. H.; Luo, Q.; Guo, M.; Li, Q. S. What are the roles of N3 and N5 rings in designing polynitrogen molecules? *Dalton T* 2012, *41*, 12075.
(15)    Zahariev, F.; Hu, A.; Hooper, J.; Zhang, F.; Woo, T. Layered single-bonded nonmolecular phase of nitrogen from first-principles simulation. *Phys Rev B* 2005, *72*, 214108.
(16)    Ceppatelli, M.; Pagliai, M.; Bini, R.; Jodl, H. J. High-Pressure Photoinduced Synthesis of Polynitrogen in δ and ϵ Nitrogen Crystals Substitutionally Doped with CO. *The Journal of Physical Chemistry C* 2015, *119*, 130-140.
(17)    Caracas, R.; Hemley, R. J. New structures of dense nitrogen: Pathways to the polymeric phase. *Chem Phys Lett* 2007, *442*, 65-70.
(18)    Zahariev, F.; Dudiy, S. V.; Hooper, J.; Zhang, F.; Woo, T. K. Systematic Method to New Phases of Polymeric Nitrogen under High Pressure. *Phys Rev Lett* 2006, *97*, 155503.
(19)    Eremets, M. I.; Gavriliuk, A. G.; Trojan, I. A.; Dzivenko, D. A.; Boehler, R. Single-bonded cubic form of nitrogen. *Nat Mater* 2004, *3*, 558-563.
(20)    Cacace, F.; de Petris, G.; Troiani, A. Experimental detection of tetranitrogen. *Science* 2002, *295*, 480-481.
(21)    Huang, H.; Zhang, G.; Liang, S.; Xin, N.; Gan, L. Selective Synthesis of Fullerenol Derivatives with Terminal Alkyne and Crown Ether Addends. *The Journal of Organic Chemistry* 2012, *77*, 2456-2462.
(22)    Cacace, F. Experimental Detection of Tetranitrogen. *Science* 2002, *295*, 480-481.
(23)    Rennie, E. E.; Mayer, P. M. Confirmation of the "long-lived" tetra-nitrogen (N4) molecule using neutralization-reionization mass spectrometry and ab initio calculations. *The Journal of Chemical Physics* 2004, *120*, 10561-10578.
(24)    Glukhovtsev, M. N.; Jiao, H.; Schleyer, P. v. R. Besides N2, What Is the Most Stable Molecule Composed Only of Nitrogen Atoms?†. *Inorg Chem* 1996, *35*, 7124-7133.
(25)    Greschner, M. J.; Zhang, M.; Majumdar, A.; Liu, H. Y.; Peng, F.; Tse, J. S.; Yao, Y. S. A New Allotrope of Nitrogen as High-Energy Density Material. *J Phys Chem A* 2016, *120*, 2920-2925.
(26)    Engelke, R.; Stine, J. R. Is N8 Cubane Stable. *J. Phys. Chem.* 1990, *94*, 5689-5694.
(27)    Hirshberg, B.; Gerber, R. B.; Krylov, A. I. Calculations predict a stable molecular crystal of N8. *Nat Chem* 2013, *6*, 52-56.
(28)    Eaton, P. E.; Cole, T. W. Cubane. *J Am Chem Soc* 1964, *86*, 3157-3158.
(29)    Eaton, P. E.; Xiong, Y. S.; Gilardi, R. Systematic Substitution on the Cubane Nucleus - Synthesis and Properties of 1,3,5-Trinitrocubane and 1,3,5,7-Tetranitrocubane. *J Am Chem Soc* 1993, *115*, 10195-10202.
(30)    Eaton, P. E.; Zhang, M. X.; Gilardi, R.; Gelber, N.; Iyer, S.; Surapaneni, R. Octanitrocubane: A New Nitrocarbon. *Propellants, Explos., Pyrotech.* 2002, *27*, 1-6.
(31)    Chaban, V. V.; Prezhdo, O. V. Energy Storage in Cubane Derivatives and Their Real-Time Decomposition: Computational Molecular Dynamics and Thermodynamics. *ACS Energy Letters* 2016, *1*, 189-194.
(32)    Engelke, R. Calculated properties of the 22 carbon/nitrogen cubanoids. *The Journal of Organic Chemistry* 1992, *57*, 4841-4846.
(33)    Engelke, R. Ab initio calculations of ten carbon/nitrogen cubanoids. *J Am Chem Soc* 1993, *115*, 2961-2967.
(34)    Andersen, H. C. Molecular-Dynamics Simulations at Constant Pressure and-or Temperature. *J Chem Phys* 1980, *72*, 2384-2393.
(35)    Stewart, J. J. P. Optimization of parameters for semiempirical methods V: Modification of NDDO approximations and application to 70 elements. *J Mol Model* 2007, *13*, 1173-1213.
(36)    Stewart, J. J. P. Application of the PM6 method to modeling the solid state. *J Mol Model* 2008, *14*, 499-535.
(37)    Stewart, J. J. P. Application of the PM6 method to modeling proteins. *J Mol Model* 2009, *15*, 765-805.
(38)    Stewart, J. J. P. Optimization of parameters for semiempirical methods VI: more modifications to the NDDO approximations and re-optimization of parameters. *J Mol Model* 2013, *19*, 1-32.
(39)    Korth, M.; Pitoňák, M.; Řezáč, J.; Hobza, P. A Transferable H-Bonding Correction for Semiempirical Quantum-Chemical Methods. *J Chem Theory Comput* 2010, *6*, 344-352.

(40) Řezáč, J.; Hobza, P. Advanced Corrections of Hydrogen Bonding and Dispersion for Semiempirical Quantum Mechanical Methods. *J Chem Theory Comput* 2012, *8*, 141-151.
(41) Risthaus, T.; Grimme, S. Benchmarking of London Dispersion-Accounting Density Functional Theory Methods on Very Large Molecular Complexes. *J Chem Theory Comput* 2013, *9*, 1580-1591.
(42) Curtiss, L. A.; Redfern, P. C.; Raghavachari, K. Gaussian-4 theory. *J Chem Phys* 2007, *126*, 084108.
(43) Becke, A. D. Density-Functional Exchange-Energy Approximation with Correct Asymptotic-Behavior. *Phys Rev A* 1988, *38*, 3098-3100.
(44) Lee, C. T.; Yang, W. T.; Parr, R. G. Development of the Colle-Salvetti Correlation-Energy Formula into a Functional of the Electron-Density. *Phys Rev B* 1988, *37*, 785-789.
(45) Head-Gordon, M.; Pople, J. A.; Frisch, M. J. MP2 Energy Evaluation by Direct Methods. *Chem Phys Lett* 1988, *153*, 503-506.
(46) Allouche, A. R. Gabedit-A Graphical User Interface for Computational Chemistry Softwares. *J. Comput. Chem.* 2011, *32*, 174-182.
(47) Humphrey, W.; Dalke, A.; Schulten, K. VMD: Visual molecular dynamics. *J. Mol. Graphics* 1996, *14*, 33-38.
(48) Schmidt, M. W.; Baldridge, K. K.; Boatz, J. A.; Elbert, S. T.; Gordon, M. S.; Jensen, J. H.; Koseki, S.; Matsunaga, N.; Nguyen, K. A.; Su, S.et al. General atomic and molecular electronic structure system. *J. Comput. Chem.* 1993, *14*, 1347-1363.
(49) Maslov, M. M.; Lobanov, D. A.; Podlivaev, A. I.; Openov, L. A. Thermal Stability of Cubane C8h8. *Phys. Solid State* 2009, *51*, 645.